\documentclass[12pt,fleqn]{article} 
\usepackage{amsmath,wasysym}
\usepackage{xcolor}
  
\usepackage[shortlabels]{enumitem}
  \setlist{nosep}
  \setlist[enumerate,1]{leftmargin=20pt}
  \setlist[itemize,1]{leftmargin=10pt}
  \setlist[description,1]{leftmargin=0pt,topsep=2pt}
\usepackage[text={5.5in,9in},centering]{geometry}
\usepackage{graphicx}
\usepackage{hyperref}
\usepackage[vskip=0pt,leftmargin=24pt]{quoting}

\newcommand\Ar{\smallskip\noindent\texttt{A:\ }}
\newcommand{\Qn}{\smallskip\noindent\texttt{Q:\ }}

\title{Basic interactive algorithms: Preview}
\author{Yuri Gurevich\\
\normalsize University of Michigan, Ann Arbor, MI, USA}
\date{}
\begin{document}
\maketitle
\thispagestyle{empty}

\footnotetext{Partially supported by the US Army Research Office under  W911NF25-1-0046.}

\vspace{-20pt}
\begin{center}
\includegraphics[scale=.12]{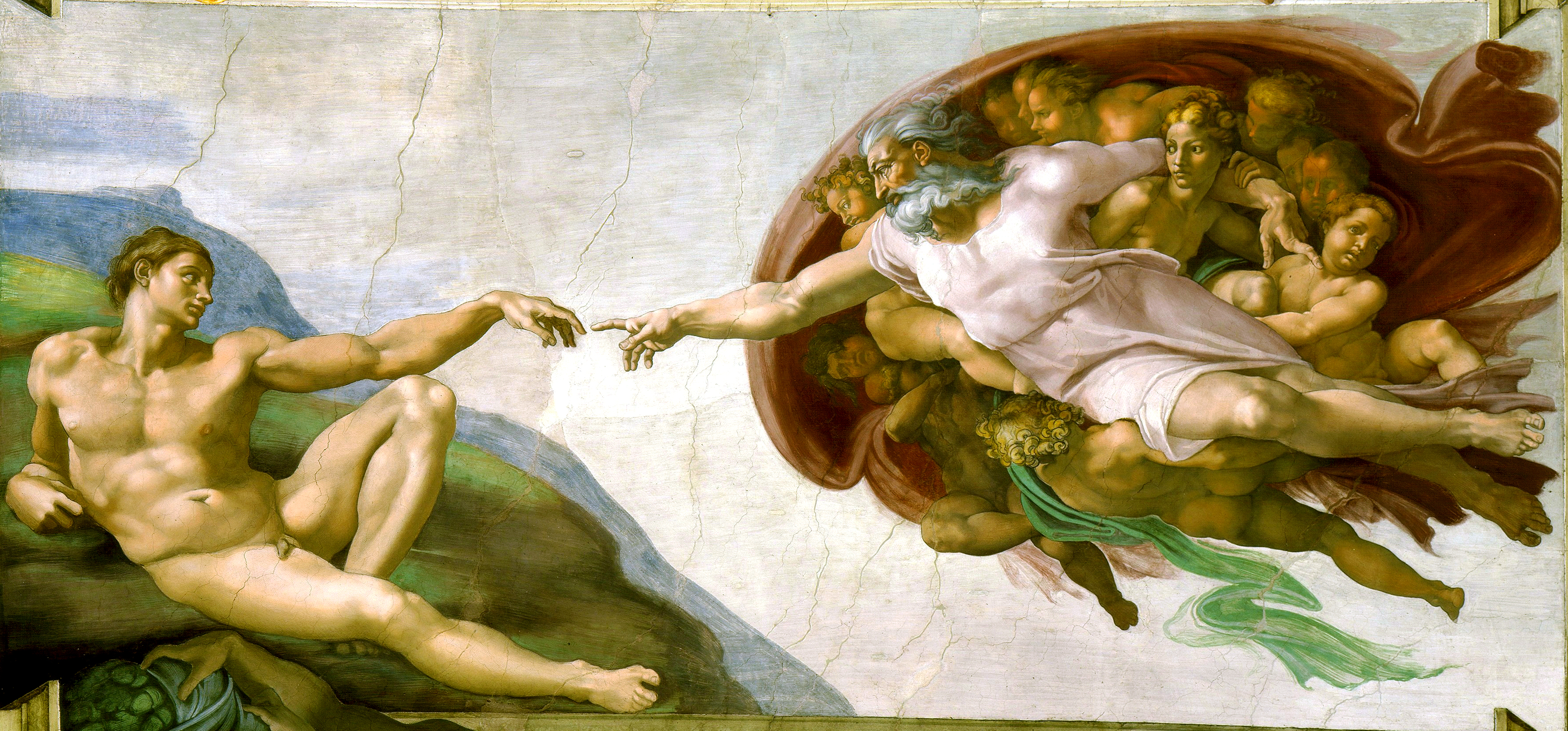} \\
{\footnotesize The power of interaction}
\end{center}

\begin{abstract}
This dialog paper offers a preview and provides a foretaste of an upcoming work on the axiomatization of basic interactive algorithms. 

The modern notion of algorithm was elucidated in the 1930s--1950s. 
It was axiomatized a quarter of a century ago as the notion of ``sequential algorithm'' or ``classical algorithm''; we prefer to call it ``basic algorithm" now. The axiomatization was used to show that for every basic algorithm there is a behaviorally equivalent abstract state machine. 
It was also used to prove the Church-Turing thesis as it has been understood by the logicians.

Starting from the 1960s, the notion of algorithm has expanded --- probabilistic algorithms, quantum algorithms, etc. --- prompting introduction of a much more ambitious version of the Church-Turing thesis commonly known as the ``physical thesis.''
We emphasize the difference between the two versions of the Church-Turing thesis and illustrate how nondeterministic and probabilistic algorithms can be viewed as basic algorithms with appropriate oracles.
The same view applies to quantum circuit algorithms and many other classes of algorithms.
\end{abstract}


\newpage
\section{Is coin flipping algorithmic?}

\Qn Consider flipping a coin. Is it algorithmic? 

\Ar Why do you ask?

\Qn People speak about randomized algorithms, involving probability distributions.  
These distributions stem from physical processes, like flipping a coin.
So a question arises: \emph{who} flips the coin?
I don't see how an algorithm could achieve that. 
An external agent has to perform the flip.

\Ar You could also ask whether a quantum measurement is algorithmic. 

\Qn Isn’t that essentially the same question? A quantum measurement also yields a probability distribution.

\Ar Well, if you doubt that an algorithm can flip a coin, you might be even more skeptical about an algorithm performing a quantum measurement. 
That process involves Mother Nature, after all.

\Qn But coin flipping involves Mother Nature as well, doesn't it?

\Ar You are right. 
The difference is that quantum measurement involves an aspect of nature that we are not accustomed to and don't fully understand, whereas coin flips have been well understood for a long time. 

In any case, I stick to the traditional view that algorithms are inherently deterministic, making ``nondeterministic algorithm'' a contradiction in terms. 
Yogi Berra, an American philosopher and baseball player, once illustrated this: ``When you come to a fork in the road, take it." 

\Qn How do you reconcile this view with the widespread use of the term ``nondeterministic algorithm''?

\Ar This could be just a figure of speech.
Nondeterministic algorithms can be viewed as deterministic algorithms that interact with their environment where someone makes the necessary choices, possibly by flipping a coin. 
Some authors say that a random sequence of 0's and 1's is part of the input, so that the necessary choices are made ahead of time.

Alternatively --- and quite legitimately --- one can broaden the notion of algorithm, just as the notion of numbers was broadened.
From positive integers all the way to real numbers, then to complex numbers and beyond. 

\Qn Let's consider a simple example of a nondeterministic algorithm. 

\Ar Here's a classic ruler-and-compass algorithm with minimal nondeterminism. The setting is a fixed Euclidean plane.
Given a circle $C$, its center $p$, and a point $q$ outside of $C$, the algorithm constructs a tangent from $q$ to $C$. 
\setlength{\mathindent}{0pt}
\begin{equation}\label{tangent}
\begin{aligned}
&1.\ \texttt{draw the midpoint $r$ between $p$ and $q$};\\
&2.\ \texttt{draw the circle $D$ centered at $r$ and passing through $q$};\\
&3.\ \texttt{choose a point $s$ where the circles $C,D$ intersect};\\
&4.\ \texttt{draw the line through $q$ and $s$}.
\end{aligned}
\end{equation}
\setlength{\mathindent}{\parindent}

\noindent
The resulting line through $q$ and $s$ is the desired tangent to $C$.

\includegraphics[scale=.5,trim=225pt 150pt 0 120pt,clip] 
{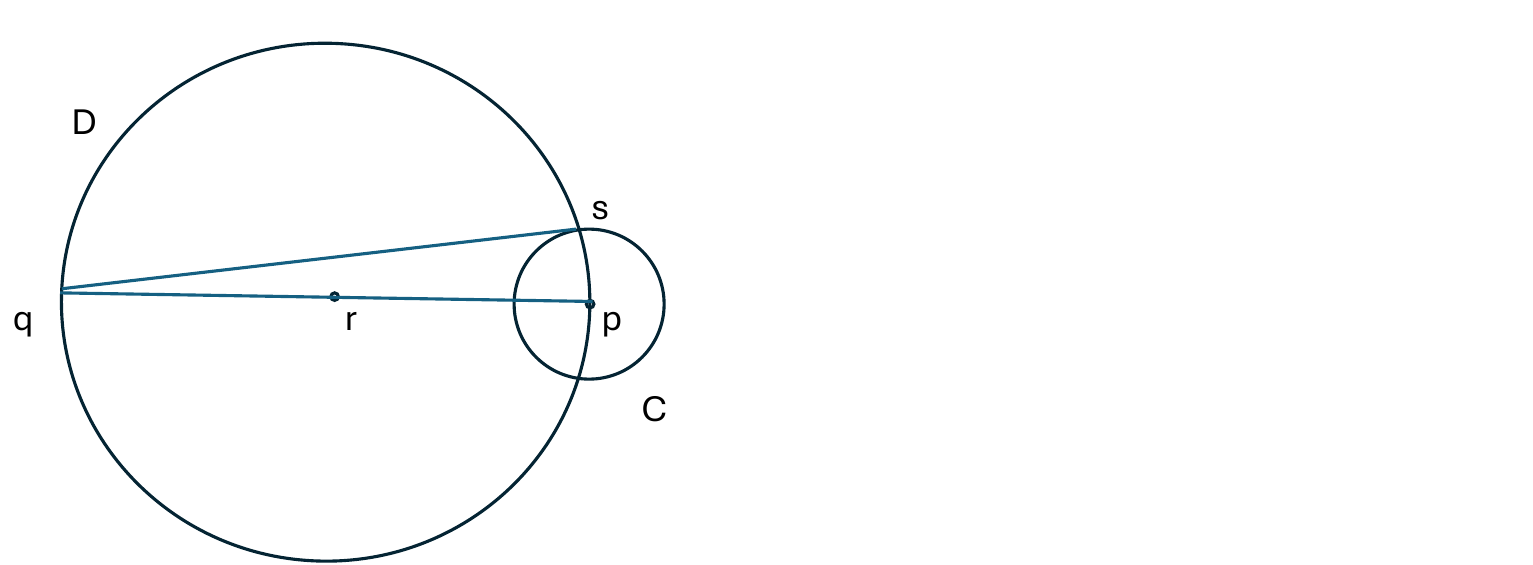} 

Instruction~3 involves a nondeterministic choice.
Either intersection point will work, but the choice must be made.
The algorithm doesn't specify which one to choose; presumably, that’s left to the executor. Aside from this step, the construction is fully deterministic. 

\section{Two vastly different theses}

\Qn Before we plunge into interactive algorithms, let me ask you, again, about the Dershowitz-Gurevich derivation of the Church-Turing thesis in \cite{G188}.
Peter Shor is critical of it, but I'm interested in exploring the concerns he raises.
\begin{quoting}
``The Dershowitz-Gurevich paper says nothing about probabilistic or quantum computation. It does write down a set of axioms about computation, and prove the Church-Turing thesis assuming those axioms. However, we're left with justifying these axioms. Neither probabilistic nor quantum computation is covered by these axioms (they admit this for probabilistic computation, and do not mention quantum computation at all), so it's quite clear to me these axioms are actually false in the real world, even though the Church-Turing thesis is probably true'' 
\cite{StackExchange}.
\end{quoting}
But let me start with this: Do you think he read the paper?

\Ar I'd guess that he just skimmed the abstract and then searched the text for ``probabilistic" and ``quantum."

\Qn Is this arrogance or misperception?

\Ar Both, I think. 
The misperception would disappear if he read just a bit beyond the abstract.

There are two vastly different interpretations of the Church-Turing thesis in play.
The thesis may be formulated thus: 
\begin{quote}
\texttt{Church-Turing thesis.} \emph{Every effectively calculable string-to-string function is Turing computable.}
\end{quote}
A question arises: what does ``effectively calculable'' mean?

One interpretation is traditional in logic.
In the 1930s -- 1950s, the meaning of ``effectively calculable'' was not in dispute.
Logicians, including Church and Turing, --- Turing wrote his dissertation in logic, under Church --- had robust intuition about it.
This intuition is elucidated in many books, for example in the influential books \cite[\S62]{Kleene52}, \cite[\S1.1]{Rogers}, and \cite[\S9]{Shoenfield}. 
In this interpretation, effective calculation was deterministic.
For example Turing writes: ``The behaviour of the computer at any moment is determined by the symbols which he is observing, and his `state of mind' at that moment'' \cite[\S9]{Turing}.
Let me call this original interpretation of the thesis ``classical'' (though we should be careful with this term because, in quantum computing, ``classical''  means ``not quantum'').

The other interpretation is also natural in a sense, especially if you don't know the history of the subject.
``Effectively calculable'' can be interpreted  as physically computable, which allows  probabilistic and quantum computations as well as highly parallel,  distributed, etc.
This broader interpretation is commonly known as the ``physical thesis.''

\Qn I see. You derive the classical thesis, and you think that Shor has in mind the physical thesis.

\Ar Yes, that is what I think.
\S1 of our paper is about effectivity.
If Shor read just a little beyond the abstract, he would know that we prove the classical thesis.

\Qn Why is it ``quite clear'' to Shor that your axioms ``are actually false in the real world?''

\Ar I can only speculate about that.
The axioms fail to cover probabilistic and quantum algorithms, so they are wrong and the proof does not establish the thesis.
Nevertheless, the thesis ``is probably true.''
Of course I think that it cannot be true \cite{G241}.

\Qn The speculation seems reasonable.
But if the axioms imply the physical thesis then the thesis must have been  formulated in mathematical terms, which is of independent interest.
That should have occurred to Shor unless indeed he spent infinitesimal time on your paper.
Anyway, is there a convincing formulation of the physical thesis?

\Ar I don't think so.
Early on, Robin Gandy attempted such a formulation \cite{Gandy} but the attempt was not successful. In particular, it didn't cover distributed computing.

\Qn Are your axioms true in the real world?

\Ar A better question is what algorithms satisfy the axioms?

\Qn Do you admit that your axioms don't cover probabilistic computations?

\Ar ``Admit'' is the wrong word. 
Probabilistic algorithms are out of scope of our paper because they aren't classical.
The relevant paragraph says:
``Methods satisfying the Sequential Postulates include ...
On the other hand, the postulates exclude ...
They are also meant to exclude nondeterministic methods ..., probabilistic methods (like Rabin’s algorithm
for testing primality), ..."
But the same paragraph includes this footnote: 
\begin{quoting}
``Large classes of such non-classical algorithms are covered by the generalizations of the ASM Theorem in \cite{G157,G166/170/171,G176/182,GR}.''
\end{quoting}

\Qn I know from our earlier conversations that ``ASM'' stands for ``abstract state machine.''
But what do you mean by the ASM theorem?

\Ar Let's take a quick dive into the history of the ASM project.
It started in mid 1980s as a computer theory project --- I wanted to understand what algorithms are.
That led me to the notion of abstract state machines, originally called evolving algebras, and to the 
\begin{quote}
\texttt{ASM thesis.} \emph{For every algorithm there is a behaviorally equivalent ASM.}
\end{quote}
which I see as a natural extension of Turing's approach to the Church-Turing thesis. 
The project quickly became applied and practical.
An ASM community emerged, and ASMs were successfully used for high-level executable specifications and related tasks.
Article \cite{G098} is one example. 

The ASM thesis was first formulated for classical algorithms \cite{G098} and then extended to algorithms in general \cite{G103}.
Article \cite{G141} turned the ASM thesis, restricted to classical algorithms, called the Sequential ASM thesis in \cite{G141}, into a theorem.
That's the ASM theorem that you asked about.

\Qn Why ``sequential" rather than ``classical" thesis?

\Ar At the time, it seemed to me that the thesis can't be called classical, because it was new. Hence ``sequential'' as an imperfect substitute, emphasizing the difference from parallel, distributed, etc.
 
\Qn But the ordinary meaning of sequential algorithms does not rule out probabilistic algorithms for example.

\Ar True. It would probably have been better to call the thesis classical --- even though it was new --- because it is about classical algorithms.

\Qn Are readers ``left with justifying'' your axioms?

\Ar No.
There are four axioms (or postulates) in \cite{G188}.
The first three aim to capture all basic/classical algorithms, whether they compute a function or not.
\begin{enumerate}
\item Sequential Time Postulate  is self-evident, almost trivial.
It says that an algorithm could be viewed as (finite or infinite) automaton with states, initial states, and a transition function. 
\item Abstract State is natural, at least to logicians.
The main part of it is that states can be viewed as structures in the sense of mathematical logic.
\item Bounded Exploration Postulate says that, during a step, only a bounded part of the state is explored, namely the part given by a fixed set of expressions.
\footnote{The first two postulates occurred to me right away but it took me years to arrive at the third postulate. 
I might have arrived at it earlier if I knew at the time about Kolmogorov's insight ``An algorithmic process breaks down into separate steps of a priori bounded complexity'' \cite{Kolmogorov} in the minutes of Moscow Mathematical Society in Uspekhi Matematicheskikh Nauk, which is translated as Russian Mathematical Surveys, but minutes aren't translated.}
\end{enumerate}
Besides, the work by the ASM community provided plenty of justification.
There is little doubt that classical algorithms satisfy the three axioms. 
The surprising part is that the axioms are sufficient to capture the notion of basic/classical algorithms mathematically. 

\Qn What is the fourth axiom for?

\Ar A  string-to-string function, computed by a basic algorithm, is not necessarily Turing computable because the initial state may have too much information. 
The fourth axiom guarantees that the initial states are bare.

\section{Spec Explorer}

\Qn Do you describe ASM applications in the Dershowitz-Gurevich paper?

\Ar We probably should have, but we didn't — mainly because we submitted the paper to the Bulletin of Symbolic Logic rather than a computer science journal.

\Qn What would be a good example of ASM applications?

\Ar Let me tell you about Spec Explorer. 
The story illustrates how theoretical work on ASMs had profound practical impact.

In 1998, Microsoft Research (MSR) invited me  to create a group and apply the ASM method. 
My first hire was Wolfram Schulte, one of the most talented people in  the ASM community in Germany. 
Unlike me, he had industrial experience. 
We built a wonderful Foundations of Software Engineering group.
In a few years the group built an ASM-based tool, Spec Explorer that allowed us to write high-level executable specifications and test them against programs.
Upper management, including and especially Bill Gates, liked the tool.
But it seemed impossible to get the developers to use the tool because it required nontrivial training.
Spec Explorer was an advanced but niche tool, rich in features, beloved by testers and those who appreciated formal methods.

Then the European Union came to our rescue \smiley{}.
In the early 2000s, the EU reprimanded Microsoft—and for good reason. While outside developers were confined to official Windows interfaces, Microsoft’s own products could tap into undocumented internal protocols that offered privileged access to the core of the operating system. This asymmetry made meaningful competition on the Windows platform nearly impossible%
\footnote{The European Commission's antitrust case against Microsoft culminated in a 2004 decision (Case T-201/04).}.

The EU demanded change. Specifically, it required Microsoft to produce \emph{high-level, executable} specifications of the internal Windows protocols that its own products used. 
Microsoft lacked comprehensive, precise, high-level executable specifications. Word documents and ad hoc specs weren't sufficient.

This created a moment of crisis.
And then, unexpectedly, Spec Explorer was thrust into the spotlight.  
It could produce high-level executable specs using state machines and C\# annotations.
It could generate test suites and simulate behaviors.
It was suitable for model checking and detecting inconsistencies.

The Windows Division picked up Spec Explorer to model protocols. 
The tool was ``dumbed down" a bit for broader adoption --- many industrial teams want minimal knobs.
It was also strengthened and productized to scale to the thousands of pages of protocol documentation eventually published.
It became an industrial-strength workhorse. 
Teams worked tirelessly to generate models and produce the documentation needed to satisfy the EU’s demands.

The specifications were released as part of the Microsoft Open Specifications Promise.
Microsoft paid significant fines (over €1.6 billion across multiple years).
The case marked a major shift toward transparency and openness, especially regarding protocols and APIs.

Ultimately, it was, however, a Pyrrhic victory for the EU. While Microsoft did produce thousands of pages of protocol specifications and did improve transparency, the underlying technology landscape was shifting. The relevance of those protocols was already beginning to fade. Web services, cross-platform frameworks, and cloud computing were quickly rendering the entire issue obsolete.

Still, for a brief and intense period, formal modeling and executable specifications were at the center of one of the most consequential regulatory battles in software history—and Spec Explorer played a key role in bridging the gap between legal compliance and technical execution.

\section{Examples}

\Qn You said that classical algorithms are deterministic, but the tangent algorithm is not. 
It is a ruler-and-co mpass algorithm that comes from antiquity; it can't be more classical.
\begin{center}
\includegraphics[scale=.25]{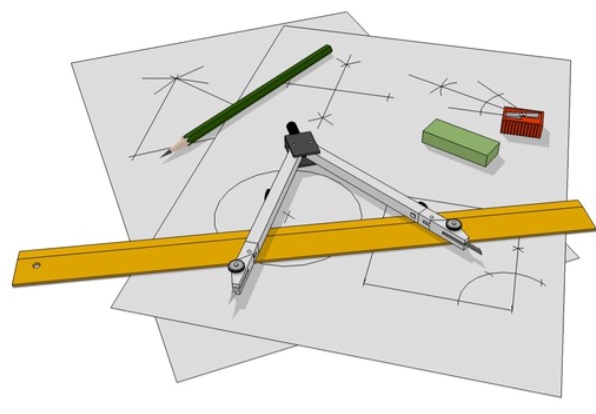} \\
{\footnotesize Credit: Wikipedia \cite{Wiki}}
\end{center}
\Ar This is a matter of definition. 
The notion of algorithm was elucidated in logic work of the 1930s--1950s. 
It is these algorithms that I called classical.
They are deterministic.
The notion of nondeterministic algorithms was formally introduced by Stephen Cook only in 1971 \cite{Cook}
(and independently by Levin  in 1973 \cite{Levin}, with a slightly different terminology), though Michael Rabin and Dana Scott introduced  nondeterministic finite automata in 1959 \cite{RS}.

By the way, in classical geometry, the tangent algorithm is  considered deterministic. 
Both solutions are equally valid; the choice is left to the user.

\Qn  I've read Church's and Turing's thesis papers, \cite{Church} and \cite{Turing},  and I don't recall them using the word ``algorithm.''

\Ar You are right. 
The term algorithm existed earlier, but it referred to calculations with Arabic numerals. 
Logicians spoke about ``effective method'', ``mechanical procedure'', and ``rule of calculation.'' 
It wasn't until the 1950s–60s, with the rise of computer science, that ``algorithm'' took on its current meaning: a finite, unambiguous, effective procedure for solving a problem.

\Qn What ASM is behaviorally equivalent to the tangent algorithm? 

\Ar Let me start from afar.
Often, especially in small examples, the form of ASMs from \cite{G141} is used.
It is an implicit iteration of a generic one-step rule $R$ and is given just by $R$.
The most common convention is that $R$ is executed repeatedly until, if ever, you reach a fixed-point state $X$, so that, in state $X$, executing $R$  leaves the state unchanged.

This convention is not appropriate for interactive algorithms. Executing $R$ in a fixed-point state isn't innocent if $R$ involves any interaction with the environment because this interaction is visible externally.
Fortunately, the three axioms of \cite{G141} imply that the halting condition of a basic algorithm can be expressed by a Boolean-valued term. 
Accordingly, the program of a basic --- in the sense of basic algorithms -- ASM can be given in the form 
\begin{equation*}
\begin{aligned}
&\texttt{do until }H\\
&\quad R
\end{aligned}
\end{equation*}
where $R$ is a generic step rule and $H$ a halting condition.
For example, consider the following version of Euclid's algorithm for computing the greatest common divisor $d = GCD(a,b)$ of two nonnegative integers $a\ge b$.
\begin{equation}\label{euclid}
\begin{aligned}
&\texttt{while $b>0$ do}\\
&\quad(a:=b) \parallel (b:= a \bmod  b);\\ 
&d := a
\end{aligned}
\end{equation}
In this case, the program of a simulating ASM could be
\begin{equation*}
\begin{aligned}
&\texttt{do until } d=a\\
&\quad\texttt{if\ $b>0$\ then\ $(a:=b) \parallel (b:= a \bmod  b)$}\\
&\quad\texttt{else $d:=a$}
\end{aligned}
\end{equation*}
where it is presumed that $d$ is initially undefined (or has a value, like $-1$, that $a$ cannot possibly have).

\Qn Typically, when you convert a while loop into a do-until loop, you just negate the condition. 

\Ar Since we insist that the do-until loop is the whole program, it needs to ``suck in'' the assignment $d:=a$ and thus becomes one step longer.

\Qn Implicit iteration makes good sense in this case because the step rule makes it clear when to stop

\Ar Indeed, if a program (i)~merely computes the value of a term $t$ that involves no oracle queries and (ii)~at the final step executes an assignment $o := t$ where $o$ is
an output variable, then the obvious halting condition is $o = t$
and the do-until loop can be made implicit.
In the case of Euclid's algorithm, the program of the simulating ASM can be given by the generic step rule
\begin{equation*}
\begin{aligned}
&\quad\texttt{if\ $b>0$\ then\ $(a:=b) \parallel (b:= a \bmod  b)$}\\
&\quad\texttt{else $d:=a$}
\end{aligned}
\end{equation*}
\Qn Let's return to the tangent algorithm. 
I'd expect that the program of an ASM that naturally  simulates \eqref{tangent} would be similar to \eqref{tangent}.

\Ar This is true in a sense, but the only ASM  commands used to form a generic step rule in \cite{G141}  are assignments, conditionals, and parallel compositions.
Sequentiality within a step is handled by conditionals and sequentiality of different steps by iteration.

To deal with geometric objects, we use a restricted incidence relation Inc of type Point $\times$ Circle $\to$ Boolean, and we need some program variables and function symbols:
\begin{itemize}
\item[-] program variables $r,s$ of type {\tt Point}, $D$ of type {\tt Circle}, and $T$ of type {\tt Line}, 
\item[-] binary function symbols $Cl, L,M$ with domain type {\tt Point $\times$ Point} and range types {\tt Circle, Line}, and {\tt Point} respectively, and 
\item[-] a binary function symbol $I$ of domain type {\tt Circle $\times$ Circle} and range type {\tt Point}.
\end{itemize}
The intended meaning of the binary functions is as follows. 
Let $x,y$ be distinct points and $A,B$ distinct intersecting circles. 
Then
\begin{itemize}
\item[-] $Cl(x,y)$ is the circle through $y$ with center $x$.
\item[-] $L(x,y)$ is the line through $x$ and $y$. 
\item[-]$M(x,y)$ is the midpoint between $x$ and $y$.
\item[-] $I$ is an oracle function. $I(A,B)$ is a point in the intersection of the two circles.
\end{itemize}

Making the do-until loop implicit, we can mimic the structure of \eqref{tangent}.
\begin{equation}\label{tangent2}
\begin{aligned}
& r := M(p,q) \parallel \\
&\texttt{if $r = M(p,q)$ then  } D := C(r,q) \parallel \\
&\texttt{if $D = C(r,q)$ then } s := I(C,D) \parallel \\
&\texttt{if $\text{Inc}(s,C)\land \text{Inc}(s,D)$ then } T := L(q,s).
\end{aligned}
\end{equation}

\Qn Why isn't the guard $s = I(C,D)$ in the last line?

\Ar Because we cannot guarantee that the second oracle call $I(C,D)$ will give the same result as the previous one.
(In \cite{G166/170/171}, we have a convention that, within a step, the same queries have the same anwer.)

By the way, in the case of the tangent algorithm, we can do better:
\begin{flalign*}
& T:= L(q,I(A,C(M(p,q),q))).
\end{flalign*}

\Qn I'd like to see an example where you start with a probabilistic algorithm, say with the Rabin Primality Test.

\Ar Ok, let me first recall what it is all about.

\noindent
\texttt{Problem:\ } Given an integer $n$ determine with sufficiently high probability  whether $n$ is prime.

\noindent
\texttt{Algorithm:\ } Let $a,i$ be integer variables initialized to 1, $k$ a positive integer constant, \texttt{prime} a Boolean variable initialized to \texttt{true}, and \texttt{Random}  a probabilistic oracle that, given integers $b<c$, selects a number in the segment $[b,c]$ according to the uniform probability distribution.

\begin{align*}
&\texttt{do until prime = false or $i > k$}\\
&\quad\texttt{choose a random integer $a$ such that $2 \le a \le n-2$};\\
&\quad\texttt{if $a^{n-1} \ne 1$ then prime :=  false}\\
&\quad\texttt{else increment $i$ by 1} 
\end{align*}

The probability of a false positive --- $n$ is composite but \texttt{prime} retains value \texttt{true} --- decreases exponentially with $k$.
So the desired precision can be given by the value of $k$.

\noindent
\texttt{ASM. } The program of a simulating ASM could be
\begin{align*}
&  \texttt{do until }(\texttt{prime = false}) \lor (i > k)\\
&\quad a := \texttt{Random}(2,n-2) \parallel i := i+1\\
&\quad \texttt{if $a^{n-1}\bmod n \ne 1$ then prime:=false}
\end{align*}

\Qn You said that basic algorithms are not parallel, and you keep using parallel composition in the examples.

\Ar In \cite{G141}, parallelism is subject to the following   constraints.
\begin{itemize}
\item It is used only within a step, and
\item the number of components is a priori bounded.
\end{itemize}

\Qn I thought that a single step results in at most one state changing action. 
Parallel actions within a single step seem to contradict that.

\Ar Well, even Turing machines permit these two parallel actions during a single step: changing the control state and moving the head.


\section{On basic interactive  algorithms}

\Qn You say that nondeterministic algorithms are naturally interactive. 
Deterministic algorithms also could be interactive.

\Ar Deterministic algorithms are typically interactive, though interaction is often implicit.
For example, algorithm~\eqref{euclid} doesn't have code for computing the modulo function and therefore interaction is necessary.
But interaction is more problematic in the case of nondeterministic algorithms, as we saw above when we discussed \eqref{tangent2}.

\Qn Have you tried to extend the formalization in \cite{G141} to allow algorithms that query possibly-nondeterministic oracles?

\Ar We did.
The results were reported in lengthy publications 
\cite{G166/170/171} and \cite{G176/182}, probably too lengthy and involved for much of the intended audience.

\Qn Why so lengthy?

\Ar Those papers included detailed motivations, discussions of alternatives, explanations, detailed proofs. Also, they explicitly allowed viewing the operating system as part of the environment and viewing our algorithm as an agent in a distributed computation.

Anyway, currently Andreas Blass and I are finishing a much shorter version of \cite{G166/170/171} with many improvements and simplifications.
The axioms are the three axioms of \cite{G141} somewhat expanded.

\Qn  Give me a simple example of axiom expansion. 

\Ar The simplest example is related to the Sequential Time Postulate.

\Qn I guess you just replace the transition function with a transition relation?

\Ar This would not be enough. 
There may be steps of the algorithm that transform a state $X$ to a state $X'$ but have different interactions with the oracles.
The steps are different but a transition relation would not distinguish them.

\Qn I see. Interaction is, at least in principle, externally observable, and so the two steps exhibit different behavior.

\Ar Correct.

\Qn Does the basic interactive framework support quantum computing?

\Ar It was in fact our work on quantum circuit algorithms \cite{G254} that prompted the revision of \cite{G166/170/171}. 
Quantum circuits have quantum-transformation gates (unitary gates and measurement gates) but no code to perform them.
Accordingly, we need oracles for those quantum transformations. 
Quantum circuit algorithms are naturally basic interactive algorithms.

\subsection*{Acknowledgments}
Many thanks to Andreas Blass for commenting on the drafts of this paper so many times that he may now know the paper better than I do.

\end{document}